 \documentclass[final,5p,times,twocolumn]{elsarticle}

\usepackage{amssymb}
\usepackage{amsthm}

\usepackage{times}
\usepackage{mathptmx}
\usepackage{amsmath}
\usepackage{txfonts}
\usepackage{bm}
\usepackage{graphicx}
\usepackage{tabularx}
\usepackage{natbib}
\usepackage{hyperref}

\journal{Solid State Communications}

\begin{document}

\begin{frontmatter}

\title{Temperature dependence of universal conductance fluctuation due to development of weak localization in graphene}

\author[HCM]{D. Terasawa\corref{cor1}}
\ead{terasawa@hyo-med.ac.jp}

\author[HCM]{A. Fukuda}

\author[OIT]{A. Fujimoto}

\author[UTokushima]{Y. Ohno}

\author[ISIR]{K. Matsumoto}
\address[HCM]{Department of physics, Hyogo College of Medicine, Nishinomiya 663-8501, Japan}
\address[OIT]{Applied Physics, Faculty of Engineering, Osaka Institute of Technology, Osaka 535-8585, Japan}
\address[UTokushima]{Graduate School of Science and Technology, Tokushima University,  Tokushima 770-8501, Japan}
\address[ISIR]{The Institute of Scientific and Industrial Research, Osaka University, Ibaraki 567-0047, Japan}
\cortext[cor1]{Corresponding author}

\begin{abstract}
The temperature effect of quantum interference on resistivity is examined in monolayer graphene, 
with experimental results showing that the amplitude of the conductance fluctuation increases as temperature decreases. 
We find that this behavior can be attributed to the decrease in the inelastic scattering (dephasing) rate, which enhances the weak localization (WL) correction to resistivity.
Following a previous report that explained the relationship between the universal conductance fluctuation (UCF) and WL regarding the gate voltage dependence (D. Terasawa, {\it et al}., Phys. Rev. B {\bf 95} 125427 (2017)),  we propose that the temperature dependence of the UCF in monolayer graphene can be interpreted by the WL theory.
 \end{abstract}



\end{frontmatter}

\section{Introduction}

The quantum interference (QI) effect is of fundamental importance because it is a manifestation of wave-particle duality.
As a prominent aspect, QI effects of charged carriers are frequently observed to appear in conductance, such as Anderson localization\,\cite{Anderson}, Aharonov-Bohm effect\,\cite{ABeffect}, universal conductance fluctuation (UCF)\,\cite{LeeStone,LeeStoneFukuyama}, and weak localization (WL)\,\cite{Hikami,Bergmann}. 
Regarding the latter two QI corrections to conductance, graphene shows unconventional results\,\cite{Tikhonenko,Tikhonenko2,Ojeda-Aristizabal,Bohra}, owing to the chirality of the carriers at two inequivalent points in the graphene momentum space (valleys),  {\bf K} and {\bf K'}\,\cite{SuzuuraAndo,McCann}.
There has been much discussion on their relationship\,\cite{Horsell,KechedzhiUCF,Chuang}, and it is known that the inelastic dephasing length obtained separately from both UCF and WL analysis are in agreement\,\cite{Westervelt1,Westervelt2,YungFuChen}.
Recently, an essential development in understanding the relationship between UCF and WL has been made:
the cause of the UCF can be attributed to WL\,\cite{GrWL} in monolayer graphene.
From this reference, the ratio of inelastic scattering (dephasing) time $\tau_\varphi$ to intervalley scattering time $\tau_i$, $\tau_\varphi/\tau_i$, which causes the WL correction to conductance, varies with the UCF.
However, the effect of temperature $T$ on the relationship between the UCF and WL remains unclear.
In this communication, we demonstrate that the effect of temerature on the UCF is well described by the WL theory proposed by McCann {\it et al}.\,\cite{McCann}.
Through WL analysis, we find that $\tau_\varphi$ increases as $T$ decreases, resulting in the QI correction in the resistivity from the WL effect to become more prominent and increase the magnitude of UCF.

\section{Sample and Method}

A graphene sample is obtained using a mechanical exfoliation method \cite{Novoselov_science} on a SiO$_2$ surface that is separated by 300\,nm from an $n^+$-doped Si substrate.
We choose a monolayer flake from kish graphite of approximately 6.2 $\times$ 1.4 $\mu$m$^2$ area. The contact pattern is drawn using electron beam lithography at opposite edges of the sample (see Fig. \ref{fig1} (a)), which was also used in the previous experiment\,\cite{GrWL}.
Ohmic contact materials (10-nm-thick Pd and 100-nm-thick Au) are deposited through thermal evaporation, followed by a liftoff process in warm 1-methyl-2-pyrrolidone.
The carrier densities are controlled by varying the back gate voltage, $V_{\rm g}$, which is applied between the graphene sheet and the substrate, in accordance with the relation $dn/dV_{\rm g} = 7.2 \times 10^{10}$\,cm$^{-2}$V$^{-1}$.
The sample is first annealed at 700\,K in an H$_2$ atmosphere for 30 min,
and then placed in a sample cell containing a resistance heater.
The sample is annealed again {\it in situ}, using the heater  
for 2\,h at $\sim 410$\,K before cooling.
This allows to desorb molecules on the graphene surface, and shifts the charge neutral point (CNP)  to $V_{\rm g} \sim 0$\,V.
However, this can induce further atomic defects in the sample due to the amorphous carbon produced from residual hydrocarbons during the annealing\,\cite{JinpyoHong}.
Figure \ref{fig1} (b) shows the data of the Raman shift spectrum for the sample.
The number of layers was confirmed by analysis of this spectrum\,\cite{Ferrari200747}, with the prominent $D$ peak at 1320\,cm$^{-1}$ indicating a large number of atomic defects.
The resistances are measured using a standard AC lock-in technique with a frequency of 37\,Hz.
To prevent the self-heating effect caused by the carriers, we use a 10-nA\,r.m.s. source-drain current (current density: $7.1 \times 10^{-3}$\,Am$^{-1}$).
A He-free Gifford-McMahon refrigerator is used to cool both the sample and a superconductor magnet with a maximum magnetic field of 8\,T.

\section{Results and Discussion}

\begin{figure}  [t]
\includegraphics[width=1\linewidth]{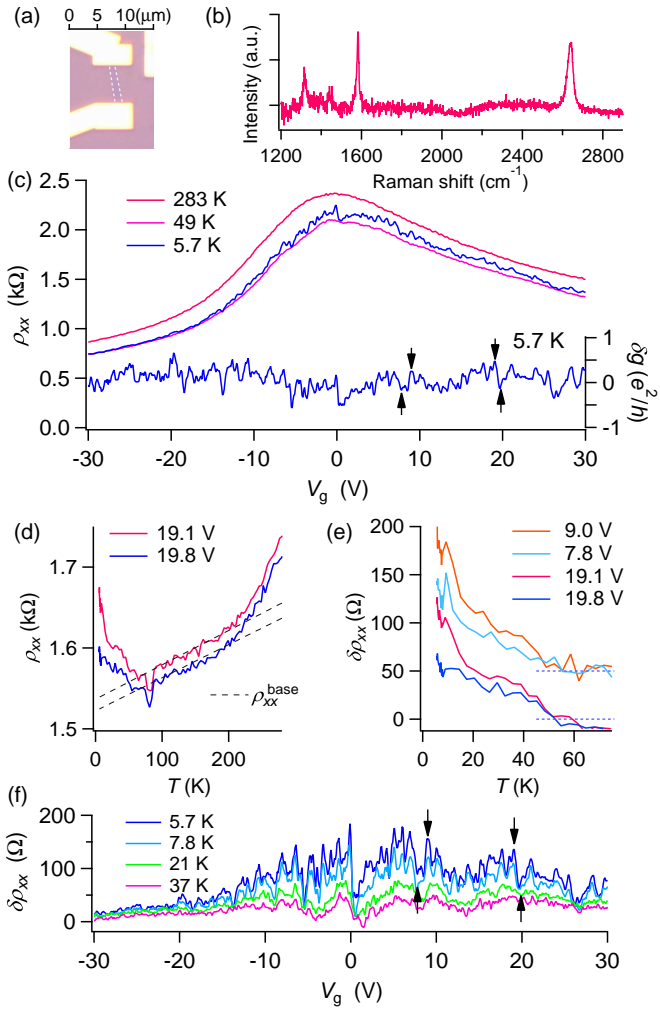}
\caption{\label{fig1}(Color online) (a)\,Optical image of the examined sample. (b)\,Raman spectrum at 632.8\,nm for the examined sample. (c)\,$\rho_{xx}$ (left axis) for different $T$ values and $\delta g = \rho_{xx}/\langle \rho_{xx} \rangle^2$ (right axis) as functions of $V_{\rm g}$ at 0 T. $\langle \rho_{xx} \rangle$ is obtained by polynomial fitting $\rho_{xx}$. The arrows indicate $V_{\rm g}=7.8$, 9.0, 19.1, and 19.8\,V. (d)\,$\rho_{xx}$ as a function of $T$ for $V_{\rm g}=19.1$ and $19.8$\,V. The dashed lines represent $\rho_{xx}^{\rm base}$, which are fitted over the region in which $\rho_{xx}$ exhibits linear dependence on $T$. (e)\,$\delta\rho_{xx}$ as a function of $T$. The data are offset by $+50$\,$\Omega$ for $V_{\rm g}=7.8$, and 9.0\,V (f) $\delta \rho_{xx} $ as a function of $V_{\rm g}$ for different $T$ values.}
\end{figure}

Figure \ref{fig1} (c) shows the longitudinal resistivity, $\rho_{xx}$, as a function of $V_{\rm g}$ for several values of the temperature, $T$ (5.7, 49, and 283\,K).
The gate voltage at the CNP, $V_{\rm CNP}$, at which the carrier type changes from electrons ($V_{\rm g} > V_{\rm CNP}$) to holes ($V_{\rm g} < V_{\rm CNP}$), is 0.78\,V for this sample.
As is observed, $\rho_{xx}$ exhibits asymmetric dependence on $V_{\rm g}$ for both  the electron and hole regions, due to the invasive contact with Pd\,\cite{Huard}. Therefore, the carrier mobility, $\mu$, varies in both cases, with $\mu =1/(ne\rho_{xx}) \approx 0.34 \pm 0.13$ and $0.46 \pm 0.09 $\,m$^2$V$^{-1}$s$^{-1}$ for electrons and holes, respectively, at $V_{\rm g}$ approximately 10\,V away from the CNP.
For $-30 \leq V_{\rm g} \leq 30$\,V, we find $k_{\rm F} \ell \geq 5$, where $k_{\rm F}$ is the Fermi wave number and  $\ell$ is the mean free path from the Drude model ($\ell = h/(2e^2k_{\rm F}\rho_{xx}$)), indicating that the system is considered to be in a diffusive metal. 
We exclude the charge neutral region ($-5.2 \lesssim V_{\rm g} \lesssim 7.5$\,V), in which the transport magnetic field $B_{\rm tr}=\hbar/(e\ell^2)$ changes linearly as $V_{\rm g}$ increases, from the following analysis.
The temperature dependence of $\rho_{xx}$ shows a metallic behavior for $T \gtrsim 60$\,K, i.e., $\rho_{xx}$ decreases as $T$ decreases (see Fig.\,\ref{fig1} (d)), due to the decrease in resistivity of the contact metal.
However, $\rho_{xx}$ gradually starts to increase at approximately $T \sim 60$\,K.
In Fig. \ref{fig1} (d), typical temperature dependences of $\rho_{xx}$ are shown for $V_{\rm g} =19.1$ and $19.8$\,V, corresponding to a local maximum (LMax) and local minimum (LMin), respectively.
This figure demonstrates that the value of $\rho_{xx}$ at $V_{\rm g} = 19.1$\,V for $T < 50$\,K is greater than that at $V_{\rm g} =19.8$\,V.
In Fig.\,\ref{fig1} (e), we plot the estimated increase in the resistance, $\delta\rho_{xx}$, as a function of $T$ for $V_{\rm g} =7.8$, 9.0, 19.1, and 19.8\,V.
We use the linear region of the plot of $\rho_{xx}$ against temperature as the base $\rho_{xx}$ behavior, and extrapolate the fitted line to the lowest temperature to obtain the value of $\rho_{xx}^{\rm base}$  (the dashed lines in Fig. \ref{fig1} (d)). We then deduce the estimated $\delta\rho_{xx}$ from
\begin{equation}
\delta \rho_{xx}(T) = \rho_{xx}(T) - \rho_{xx}^{\rm base}(T).  \label{eq_drho}
\end{equation}
The  values of $\delta\rho_{xx}$ increase at $V_{\rm g}=9.0$ and 19.1\,V, where $\rho_{xx}$ reaches LMax in the fluctuation. 
The $\delta \rho_{xx}$ results for several values of $T$ are shown in Fig. \ref{fig1} (f).

In the presence of the magnetic field $B$, $\rho_{xx}$ is reduced because the WL effect is broken.
This magnetic field correction in $\rho_{xx}$ has been formulated by McCann {\it et al}.\,\cite{McCann}.
They revealed that intervalley scattering breaks the carrier chirality in graphene, and allows constructive interference in a closed scattering path.
Considering that $\Delta \rho_{xx}(B) = \rho_{xx}(B) - \rho_{xx}(0)$,  
\begin{gather}
\Delta \rho_{xx}(B) = -\frac{e^2 \rho_{xx}^2}{\pi h} \left [ F \left(\frac{B}{B_\varphi}\right)-F\left(\frac{B}{B_\varphi+2B_i}\right) - 2F\left( \frac{B}{B_\varphi + B_\ast} \right) \right ],  \label{eqDR} \\
F(z) = \ln z+\varPsi \left( \frac{1}{2} + \frac{1}{z} \right),\mspace{18mu} B_{\varphi, i, \ast} = \frac{\hbar}{4De}\tau^{-1}_{\varphi, i, \ast}. \label{eqtau}
\end{gather}
Here,  $\varPsi$ is the digamma function, $D=v_{\rm F}\ell/2$ denotes the diffusion constant with $v_{\rm F}$ denoting the Fermi velocity ($ \approx 10^6$\,m/s),
and $\tau_\varphi$ and $\tau_i$ represent the dephasing time and intervalley scattering time, respectively.
$\tau_\ast^{-1} = \tau_i ^{-1} + \tau_z^{-1} + \tau_w^{-1}$, where $\tau_z$ denotes the intravalley scattering time, and $\tau_w$ denotes the intravalley warping time combined with the chirality breaking time\,\cite{Morpurgo}.
In the equation above, the first term contributes to the localization for $\tau_\varphi > \tau_i$ and the carrier inelastic scattering determines the decrease in resistance in the presence of $B$.
The second and third terms contribute to the anti-localization.
The theory predicts $\tau_\ast \sim \tau_i$ for describing the WL behavior, except in the high-carrier-density region, where the lattice warping effect becomes significant. 
Consequently, the inelastic and elastic scattering times can be extracted by modelling the resistance behavior in a small magnetic field.

\begin{figure}  [t]
\includegraphics[width=1\linewidth]{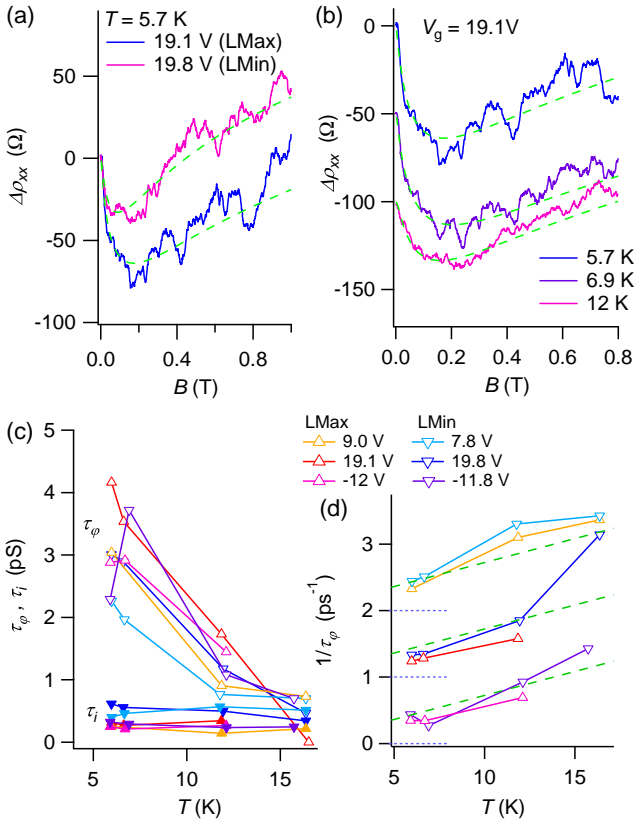}
\caption{\label{figWLT2}(Color online) (a)\,$\Delta \rho_{xx}$  as a function of $B$ for $V_{\rm g}= 19.1$\,V (LMax) and 19.8\,V (LMin) at $T=5.7$\,K. (b)\,$\Delta \rho_{xx}$ for $V_{\rm g}= 19.1$\,V as a function of $B$ for $T=$\,5.7,\, 6.9\,(offset: $-50$\,$\Omega$), and 12\,(offset: $-100$\,$\Omega$)\,K and fitting results. (c) Plots of $\tau_\varphi$ and $\tau_i$ as a function of $T$ for several values of $V_{\rm g}$. The filled triangles represent $\tau_i$ and the open triangles represent $\tau_\varphi$. (d)\,Plots of $1/\tau_\varphi$ as a function of $T$ for several values of $V_{\rm g}$ (offset: $+1$\,ps$^{-1}$ for 19.8 and 19.1\,V, $+2$\,ps$^{-1}$ for 9.0 and 7.8\,V). The dashed lines are visual guides. Marker colors are the same as for (c).}
\end{figure}

\begin{figure}  [t]
\includegraphics[width=1\linewidth]{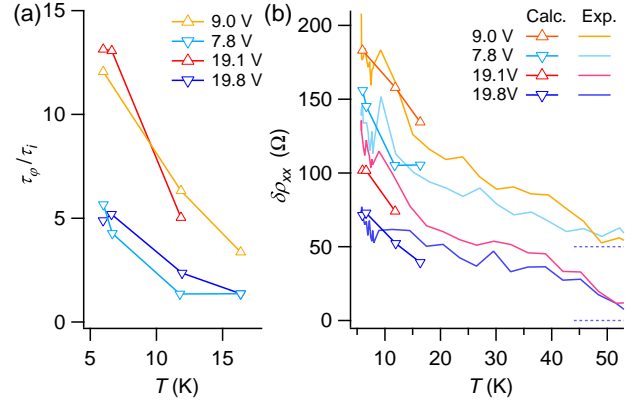}
\caption{\label{figWLT3}(Color online) (a)\,Plots of  $\tau_\varphi/\tau_i$ obtained from fitting analysis as a function of $T$. (b)\,$\delta\rho_{xx}$ derived from Eq.\,(\ref{eq_deltarhoT}) using the obtained dephasing ratio, $\tau_\varphi/\tau_i$ (upward and downward triangles). The corresponding $\delta\rho_{xx}$ obtained from Eq.\,(\ref{eq_drho}) is also shown (plain lines, identical to Fig. \ref{fig1} (e)). Data are offset by $+50$\,$\Omega$ for 9.0 and 7.8\,V.}
\end{figure}

Figure\,\ref{figWLT2}\,(a) shows $\Delta \rho_{xx}$ as a function of $B$ for LMax ($V_{\rm g}= 19.1$\,V) and LMin ($V_{\rm g}= 19.8$\,V). 
As previously demonstrated\,\cite{GrWL}, $\Delta \rho_{xx}$ is reduced by a larger amount for LMax than for LMin.
Figure\,\ref{figWLT2}\,(b) shows $\Delta \rho_{xx}$ for LMax ($V_{\rm g}= 19.1$\,V) as a function of $B$ for several values of $T$.
As $T$ increases, the amount by which $ \Delta\rho_{xx}$ is reduced decreases, indicating that $\tau_\varphi$ gradually approaches $\tau_i$ because $\tau_\varphi$ is affected by temperature-dependent scattering processes such as electron-electron scattering and electron-phonon scattering.
The scattering times are extracted from the regression analysis fit using Eq.\,(\ref{eqDR}). 
$\tau_\varphi$ and $\tau_i$ and the dephasing rate, $1/\tau_\varphi$, are shown in Fig.\,\ref{figWLT2} (c) and (d), respectively.
Detailed fitting procedures are available in Ref.\,\cite{GrWL}.
In graphene, weak electron-phonon scattering is expected\,\cite{Stauber,Hwang},
and electron-electron scattering is considered to be the major source of dephasing\,\cite{Tikhonenko,Tikhonenko2}.
Previous experiments have demonstrated that  the dephasing rate, $1/\tau_\varphi$, shows temperature dependence due to electron-electron scattering\,\cite{Tikhonenko,Tikhonenko2,Ki}.
Although the exact form of $T$ dependence of the electron-electron scattering rate is debatable\,\cite{AltshulerAronovKhmelnitsky,FukuyamaAbrahams,Narozhny}, linear-$T$ dependence is often observed in the diffusive regime ($T\ll \hbar/(k_{\rm B}\tau_{\rm m})$ )\,\cite{Tikhonenko} ($\tau_{\rm m}=\ell/v_{\rm F}$ denotes the momentum relaxation time).
For the temperature range used in our measurement,
$1/\tau_\varphi$ exhibits an approximately linear dependence on $T$ (see Fig.\,\ref{figWLT2}\,(d)),
which agrees with previous results\,\cite{Tikhonenko}.
$\tau_i$, however, is independent of $T$, and tends to be smaller for LMax than for LMin.
Theoretically, the amount by which $\rho_{xx}$ increases due to WL, $\delta \rho_{xx}$, is expressed as follows\,\cite{McCann}:
\begin{equation}
\delta\rho_{xx} = \frac{e^2 \rho^2}{\pi h}\left [ \ln\left(1+2\frac{\tau_\varphi}{\tau_i} \right) - 2\ln \left(\frac{\tau_\varphi/\tau_{\rm tr}}{1+\tau_\varphi/\tau_\ast} \right) \right].  \label{eq_deltarhoT}
\end{equation}
The second term contributes little to the change in WL, and as such, we ignore this term and conduct an analysis using $\delta\rho_{xx} \approx e^2\rho_{xx}^2/(\pi h) \cdot \ln (1+2\tau_\varphi/\tau_i )$. 
The ratio of $\tau_\varphi$ to $\tau_i$ is therefore essential to the temperature dependence of the strength of WL. 
This ratio is shown in Fig. \ref{figWLT3} (a). 
As expected, $\tau_\varphi/\tau_i$ increases as $T$ decreases, and it is 
observed that $\tau_\varphi/\tau_i$ increases more for LMax than for LMin, because $\tau_i$ tends to be smaller for LMax than for LMin.
Figure\,\ref{figWLT3} (b) shows the $\delta\rho_{xx}$ values obtained from Eq.\,(\ref{eq_deltarhoT}), along with the experimental $\delta\rho_{xx}$ results given by Eq.\,(\ref{eq_drho}) for comparison.
The results reproduce the behavior of $\delta\rho_{xx}$ very well even without any offset or coefficient parameters; therefore, Eq.\,(\ref{eq_deltarhoT}), describing WL, effectively models the temperature dependence of the UCF amplitude.

In Table \ref{tab1}, scattering lengths obtained by fitting for LMax and LMin are compared. 
Scattering lengths $L_\varphi$ and $L_i$ denote the dephasing length and the intervalley scattering length, respectively, with $L_{\varphi,i} = \sqrt{D\tau_{\varphi,i}}$.
Further comparisons are drawn with the nanographite material's grain size $L_a$ obtained from the 
empirical formula using the Raman data\,\cite{Cancado},
\begin{equation}
L_a({\rm nm}) = 2.4 \times 10^{-10}\cdot \lambda({\rm nm})^4(I_{\rm G}/I_{\rm D}).
\end{equation}
Here, $\lambda$ denotes the wavelength of the laser, and $I_{\rm G}$ and $I_{\rm D}$ denote the Raman intensity at the $G$ and $D$ peaks, respectively.
This equation yields $L_a \approx 74$\,nm for the Raman data of the sample ($I_{\rm G}/I_{\rm D} = 1.9$ and $\lambda = 632.8$\,nm), which bears a notable similarity to $L_i$.
However, $L_a$ is usually expected to match $\ell$.
This discrepancy was noted in Ref.\,\cite{Ni}, and explained in terms of the difference in energy scale when probing scattering rates between the transport and Raman measurements.
According to the analysis made in \cite{Ni}, the mean free path $\ell^\ast = \mu E_{\rm F}/(ev_{\rm F}) \approx 670$\,nm is obtained for the case of Dirac fermions, with $E_{\rm F} = 2.0$\,eV for the energies of electrons and holes generated at $\lambda = 632.8$\,nm.
This value is greater than $L_i$ and $L_a$, and therefore it is concluded that the intervalley scattering is  the transport limiting source and the dominant elastic scattering mechanism for the UCF and WL.


\begin{table} [t]
\caption{\label{tab1}Comparison of scattering lengths (nm) for LMax ($V_{\rm g}=19.1$\,V) and LMin($V_{\rm g}=19.8$\,V). }
\begin{center}
\begin{tabular}{l|cccc|c} \hline\hline
$V_{\rm g}$ &  $L_\varphi$\scriptsize{(5.7\,K)}  & $L_\varphi$\scriptsize{(12\,K)}&  $L_i$ &  $\ell$  & $L_a$ \\ \hline
 LMax (19.1\,V) &  290  & 190 & 80 & 38    &74   \\ \cline{1-5}
 LMin (19.8\,V) &  240  & 150 & 110 & 39  &  \\ \hline\hline
\end{tabular}
\end{center}
\end{table}

\section{Conclusions}

In conclusion, we have measured the effects of quantum interference correction on the electric transport phenomena in graphene.
We have observed that the UCF amplitude increases as $T$ decreases, because increasing $\tau_\varphi$ enhances the WL effect.
The temperature dependence of the UCF was reproduced by WL theory for monolayer graphene \,\cite{McCann} using the $\tau_\varphi/\tau_i$ ratio, and along with the previous report\,\cite{GrWL}, the WL theory has been shown to effectively describe the UCF behavior in graphene.

\section{Acknowledgement}

We are grateful to H. Suzuura for a fruitful discussion, to A. Sawada and H. Yayama for advice on the He -free refrigerator setup, to T. Terashima and S. Sasaki for permission to use the clean-room facilities in the LTM center of Kyoto University, and to T. Nakajima for fabricating the cryostat and electronic parts.
Funding: This work was supported by the MEXT KAKENHI on Innovative Areas (Grant No. 25103722, “Topological Quantum Phenomena”), the JSPS KAKENHI (Grants No. 24540331, No. 25870966, and No. 15K05135), Grants-in-Aid for Researchers from Hyogo College of Medicine 2011 (A.F.) and 2015 (D.T.), and a Grant for Basic Science Research Projects from the Sumitomo Foundation (Grant No. 100711). This work was performed under the Cooperative Research Program of the “Network Joint Research Center for Materials and Devices” (Grants No. 2011A10, No. 2013A18, No. 20171211, and No. 20171243).

\bibliographystyle{elsarticle-num} 


\end{document}